\documentstyle[epsf,preprint,prd,aps]{revtex}
\begin{document}
\setlength{\baselineskip}{0.6cm}
\draft
\title{Exact relation of lattice and continuum parameters \\
in three-dimensional SU(2)+Higgs theories}
\author{M. Laine\thanks{Email: mlaine@phcu.helsinki.fi}}
\address{Department of Physics, P.O. Box 9, \\
FIN-00014 University of Helsinki, Finland}
\date{April 5, 1995}
\maketitle

\begin{abstract}
\setlength{\baselineskip}{0.6cm}
The essential features of the high-temperature
electroweak phase transition are contained in
a three-dimensional super-renormalizable effective field theory.
We calculate the exact counterterms needed for lattice simulations
of the SU(2)-part of this theory. Scalar fields in both fundamental and
adjoint representations are included. The three-dimensional
U(1)+Higgs theory is also discussed.
\end{abstract}
\narrowtext

\vspace*{-14.0cm}
\noindent
\hspace*{13.0cm} \mbox{HU-TFT-95-23} \\
\hspace*{13.0cm} \mbox{hep-lat/9504001}
\vspace*{12.2cm}

\newcommand{\beq}{\begin{equation}}
\newcommand{\eeq}{\end{equation}}
\newcommand{\bea}{\begin{eqnarray}}
\newcommand{\eea}{\end{eqnarray}}
\newcommand{\eref}[1]{(\ref{#1})}
\def\undertilde#1{\mathop{\vtop{\ialign{##\cr$\textstyle{#1}$\cr%
\noalign{\kern1pt\nointerlineskip}\hfil$\mathchar"0365$\hfil\cr}}}}
\def\wideundertilde#1{\mathop{\vtop{\ialign{##\cr$\textstyle{#1}$\cr%
\noalign{\kern1pt\nointerlineskip}\hfil$\mathchar"0367$\hfil\cr}}}}

\section{Introduction}

Due to its possible
effect on the baryon number of the Universe~\cite{S1},
the cosmological electroweak phase transition should
be understood in quantitative detail. Unfortunately,
even resummed perturbation
theory~\cite{DHLLL,C,BHW,AE,BFHW,H,EEV,EQZ,FH,BFH1,FKRS1,BFH2}
may not be accurate enough, since there are infrared problems
in the ``symmetric'' high-temperature phase. This calls
for analytic~\cite{RW,S2,AM,AY,BF,BW,BP}
or lattice~\cite{EJK,BIKS,DESY,KNP,KRS,FKRS2,FKRS3}
studies of the relevant non-perturbative features.

The study of the non-perturbative features can be simplified by
combining perturbation theory and non-perturbative methods. Indeed,
the momentum scale $p\gtrsim T$ can be integrated out perturbatively,
resulting in an effective theory for length scales larger
than $1/T$. This is called dimensional
reduction~\cite{FKRS1,AP,N,L,JKP,JP,BN,J}.
The effective theory is essentially a three-dimensional (3D)
super-renormalizable SU(2) gauge theory with fundamental
and adjoint Higgs fields. The effective theory can be
studied with analytic~\cite{RW,S2,AM,AY,BF,BW,BP}
and lattice~\cite{KRS,FKRS2,FKRS3} methods with less
effort than the original four-dimensional theory.

This paper is related to lattice simulations of the
effective 3D theory. The purpose is to express the bare
parameters of the lattice action in terms of the
renormalized parameters of the effective continuum theory.
The continuum theory is regularized
in the $\overline{\rm MS}$-scheme.
The relation between lattice and continuum is needed
when results from lattice simulations are transformed
into physical values of continuum observables.
Due to the super-renormalizability of the effective theory,
the relation between lattice and continuum
can be found exactly with a two-loop calculation.
The only bare parameters having different
expressions in the two schemes are the masses of the scalar fields.
Our method is to calculate the value of a physical gauge-independent
observable in both schemes, and to compare the results.
We chose the value of the effective potential
at the minimum, apart from unphysical vacuum terms,
as the physical observable.

The relation between lattice and continuum in three-dimensional
SU(2)+Higgs theories has previously been determined in~\cite{FKRS3},
partly by analytical calculations,
and partly by lattice Monte Carlo simulations.
In the present paper, we calculate the relation fully analytically.
This should improve the accuracy of that part of the result
which was in~\cite{FKRS3} determined by lattice Monte Carlo simulations.

Let us note that our problem is analogous to the problem of
relating the values of $\Lambda_{\rm QCD}$ in $\overline{\rm MS}$ and
lattice regularization schemes in QCD~\cite{HH,DG,KNS}.
In that case, logarithmic terms arise already at
one-loop level, and there are contributions from
all orders of perturbation theory. In our case,
logarithmic terms arise only at two-loop level, and the result
is exact in the continuum limit. Hence a very high
accuracy can be reached in relating the results
of lattice simulations to continuum physics.

The theories to be discussed are the SU(2) and U(1) gauge theories
with Higgs fields in fundamental and adjoint representations.
The SU(2) theory with both fundamental and adjoint Higgs fields, or
only with a fundamental Higgs field, is relevant for the
cosmological electroweak phase
transition~\cite{FKRS1,JKP,JP}. The SU(2) theory with just an
adjoint Higgs field is relevant for studies of dimensional
reduction of the pure SU(2) gauge theory~\cite{KLMBR}.
The U(1) gauge theory with a fundamental Higgs field
could be relevant for numerical studies of
superconductivity~\cite{U1}.

The plan of the paper is the following.
In Sec.~\ref{problem}, the problem of expressing
the lattice parameters in terms of the continuum parameters
is explained, and the solution to the problem is outlined.
We follow closely~\cite{FKRS3}.
In Sec.~\ref{details}, some details of the calculation
are clarified. The results are in Sec.~\ref{results}, and the
conclusions in Sec.~\ref{concl}.
In the body of the paper, we deal with the
SU(2) + fundamental Higgs theory; results for the other
theories are collected in the Appendix.

\section{Formulation of the problem}
\label{problem}

The SU(2) + fundamental Higgs theory in three
Euclidian dimensions is described by the Lagrangian
\beq
{\cal L}  =
\frac{1}{4}F^a_{ij}F^a_{ij}+
\frac{1}{2}{\rm Tr}(D_i\Phi)^{\dagger}(D_i\Phi)+
\frac{1}{2}m_B^2{\rm Tr}\Phi^\dagger\Phi+
\frac{1}{4}\lambda_3\Bigl({\rm Tr}\Phi^\dagger\Phi\Bigr)^2.\label{Sc}
\eeq
Here $F^a_{ij}=\partial_iA_j^a-\partial_jA_i^a-g_3\epsilon^{abc}A^b_iA^c_j$,
$m_B^2=m_3^2(\mu)+\delta m_3^2$,
\beq
\Phi=\frac{1}{\sqrt{2}}(\phi_0{\bf 1}+i\phi_a\tau^a),
\label{Phi}
\eeq
and $D_i\Phi=(\partial_i+ig_3\tau^aA^a_i/2)\Phi$.
The $\tau^a$:s are the Pauli matrices,
the coupling constants $\lambda_3$ and $g_3^2$ have
the dimension GeV, and the indices take the values
$a,i,j\in\{1,2,3\}$. The sign of $g_3$ in $F^a_{ij}$ and $D_i$
is chosen so as to be compatible with~\cite{R}.

The theory defined by the Lagrangian of eq.~\eref{Sc}
is super-renormalizable~\cite{Z}. There are only two kinds of
divergences, the mass divergence and the unphysical vacuum
divergence. In the $\overline{\rm MS}$-scheme,
the exact mass counterterm needed
to make the theory finite is~\cite{FKRS1}
\beq
\delta m_3^2=-\frac{\hbar^2}{16\pi^2}\frac{\mu^{-4\epsilon}}{4\epsilon}
\biggl(\frac{51}{16}g_3^4+9\lambda_3g_3^2-12\lambda_3^2\biggr).
\label{MS3}
\eeq
Here $\hbar$ is the loop counting parameter. Due to eq.~\eref{MS3},
the renormalized mass squared is of the form
\beq
m_3^2(\mu)=\frac{\hbar^2}{16\pi^2}
\biggl(\frac{51}{16}g_3^4+9\lambda_3g_3^2-12\lambda_3^2\biggr)
\log\frac{\Lambda}{\mu}.
\label{mr}
\eeq
In the context of the electroweak phase transition, $\log\Lambda$
is known to two-loop order in terms of the physical 4D
parameters and the temperature~\cite{FKRS1}.

The two-loop vacuum counterterm
of the theory of eq.~\eref{Sc} may be chosen as
\beq
\delta V=-\frac{\hbar^2}{16\pi^2}\frac{\mu^{-4\epsilon}}{4\epsilon}
3g_3^2m_3^2(\mu).
\label{MSV}
\eeq
This choice removes the $1/\epsilon$-part from
the value of the effective potential~$V(\varphi)$
at $\varphi=0$. As a result of eq.~\eref{MSV}, $V(0)$
has an unphysical $\mu$-dependence
\beq
\mu\frac{d}{d\mu}V(0)=\frac{\hbar^2}{16\pi^2}3g_3^2m_3^2(\mu).
\label{vacmu}
\eeq
Of course, the difference between the values
of the effective potential at any two
distinct minima is free of any $\mu$-dependence.

Next, consider the theory of eq.~\eref{Sc} in lattice (L) regularization.
That is, calculations are made on a lattice with
lattice spacing $a$ and spatial extension $N$,
and in the end the limit $a\to 0, N\to\infty$ is taken.
For simplicity, in the actual momentum integrations
we take the lattice to be in the limit~$N\to\infty$ from
the beginning. The lattice Lagrangian is
\bea
{\cal L}_L & = & \frac{4}{a^4 g_3^2}\sum_{i<j}
\Bigl[1-\frac{1}{2}{\rm Tr}P_{ij}(x)\Bigr]\nonumber \\
& +  & \frac{1}{a^2}\sum_i\Bigl[
{\rm Tr}\Phi^\dagger(x)\Phi(x)-
{\rm Tr}\Phi^\dagger(x)U_i(x)\Phi(x+i)\Bigr]\nonumber \\
& + &
\frac{1}{2}m_L^2{\rm Tr}\Phi^\dagger\Phi+
\frac{1}{4}\lambda_3\Bigl({\rm Tr}\Phi^\dagger\Phi\Bigr)^2,
\label{Sl}
\eea
where
$P_{ij}(x)=U_i(x)U_j(x+i)U_i^{\dagger}(x+j)U_j^{\dagger}(x)$,
\beq
U_i(x)=\exp[\frac{i}{2}ag_3\tau^bA^b_i(x)],
\eeq
$\Phi$ is as in eq.~\eref{Phi}, and
$x+i\equiv x+ a{\bf e}_i$.
The action $S$ corresponding to the
Lagrangian in eq.~\eref{Sl} is $S=a^3\sum_x{\cal L}_L$,
where~$x$ enumerates the lattice sites.
The Lagrangian~${\cal L}_L$ is invariant under
the transformations
\beq
U_i(x)\to {U_i'}(x)=g(x)U_i(x)g^{-1}(x+i),\quad
\Phi(x)\to\Phi'(x)=g(x)\Phi(x),\label{gt}
\eeq
where $g(x)\in {\rm SU(2)}$. The path integration
over the fields $A^b_i(x)$ is defined using
the Haar measure~(see, e.g., \cite{R}),
to guarantee the gauge invariance,
and hence the renormalizability, of the theory.

In the L-scheme, the counterterms differ from
those in the $\overline{\rm MS}$-scheme.
For instance, there can be a one-loop mass counterterm in the L-scheme,
since the lattice spacing $a$ provides an extra scale that can be combined
with $g_3^2$, to make a quantity of the dimension of mass squared.
In general,
\beq
m_L^2=m_3^2(\mu)+\delta m_L^2(\hbar)+\delta m_L^2(\hbar^2) .
\label{CTs}
\eeq
As indicated by the notation, we have chosen the finite renormalized
mass squared to be exactly the same as in the $\overline{\rm MS}$-scheme,
eq.~\eref{mr}.
The bare term $m_L^2$ as a whole
is of course independent of $\mu$.

The purpose of the present paper is to express
the parameter $m_L^2$ in eq.~\eref{CTs}
in terms of the continuum
parameters~$m_3^2(\mu)$, $g_3$, $\lambda_3$,
and the lattice spacing $a$.
The one-loop mass-counterterm $\delta m_L^2(\hbar)$,
and the terms proportional to $\lambda_3g_3^2$ and $\lambda_3^2$
in the two-loop mass-counterterm $\delta m_L^2(\hbar^2)$,
have already been calculated analytically~\cite{FKRS3}.
The two-loop contribution proportional to $g_3^4$
has been computed with lattice Monte Carlo methods~\cite{FKRS3}.
Below we calculate analytically
even the contribution proportional to~$g_3^4$.

The method of calculation is the following. We extract both from
eq.~\eref{Sc} and eq.~\eref{Sl} a measurable
gauge-independent physical quantity. Since both calculations
must give the same result, $m_L^2$ can be fixed.
The simplest suitable quantity is the value of the
effective potential at the minimum, $V({\rm min})$.
To be more precise, $V(\varphi)$ contains unphysical
divergent vacuum terms, such as the one shown in eq.~\eref{MSV}.
However, apart from these, $V({\rm min})$ gives the
equation of state, and is thus physical. It has been
explicitly proved that $V({\rm min})$
is gauge-independent, when calculated consistently
in powers of $\hbar$~\cite{K,M,KLS}.

There is another, equivalent, way of formulating
the problem, without reference to the effective potential.
Indeed, one can just calculate the value of the path integral
in the broken minimum using the loop expansion.
In other words, $\phi_0$ is shifted to the classical
broken minimum, and then all the connected vacuum graphs
are calculated. It turns out that this gives just $V({\rm min})$.
To separate the vacuum terms, one should calculate the
value of the path integral in the symmetric minimum, as well.
The conceptual advantage of calculating directly
the path integral is that
complications related to fixing the gauge when
calculating the effective potential are avoided.

As a matter of fact, the problem is even simpler than
calculating $V({\rm min})$. From renormalizability, one knows that any
difference in the $\varphi$-dependent parts of $V(\varphi)$
between two schemes
could only appear in the $\varphi^2$-term. This can roughly
be seen also with simple power-counting arguments.
Indeed, any difference between two schemes arises from
the UV-region. Hence the difference should be analytic in the
parameters~$m_\varphi^2$ appearing in the propagators, which
depend quadratically on the field $\varphi$.
At one loop, the difference is then dimensionally of the form
$(1+a^2m_\varphi^2+a^4m_\varphi^4+\ldots)a^{-3}$, and
at two loops, of the form
$g_3^2(1+a^2m_\varphi^2+a^4m_\varphi^4+\ldots)a^{-2}$.
Apart form vacuum terms, higher loops give contributions
vanishing as $a\to 0$. Hence non-vanishing
differences could only arise in the~$\varphi^2$-terms
and at two-loop order. From the equation
\beq
V({\rm min})=V_0(\varphi_0)+\hbar V_1(\varphi_0)+
\hbar^2\biggl[V_2-\frac{1}{2}\frac{(V'_1)^2}{V''_0}
\biggr]_{\varphi=\varphi_0}, \label{value}
\eeq
where $\varphi_0$ is the location of the classical
broken minimum, it follows that the difference
of the $\varphi^2$-terms of
two schemes determines the difference of
the values $V({\rm min})$. In short, the counterterms
in eq.~\eref{CTs} can be fixed by requiring that
the two schemes produce the same $\varphi^2$-terms.

Let us state the problem in one more disguise:
the effective potential~$V(\varphi)$
itself is gauge-dependent, but the difference
of the effective potentials in the
L- and $\overline{\rm MS}$-schemes is not so, since it determines
the gauge-independent quantity~$m_L^2$.

To conclude this Section, we note that $V({\rm min})$ is directly
related to the measurable quantity
$\langle\frac{1}{2}{\rm Tr}\Phi^\dagger\Phi\rangle$
on lattice, by
\beq
\langle\frac{1}{2}{\rm Tr}\Phi^\dagger\Phi\rangle=
\frac{dV({\rm min})}{dm_3^2(\mu)}. \label{fdf}
\eeq
In consequence, one can actually measure the
parameter $m_L^2$ of eq.~\eref{CTs} on lattice,
by comparing lattice data to continuum perturbative results in a region
where perturbation theory works well~\cite{FKRS3}.
For such a comparison, even the mass-dependent
unphysical vacuum contributions of the type in eq.~\eref{MSV}, but
in the L-scheme, are needed,
since they enter through the right-hand side of eq.~\eref{fdf}.
Hence, we will write down also the mass-dependent
vacuum counterterms~\cite{FKRS3} below,
although mass-independent vacuum terms are neglected.

\section{Details of the calculation}
\label{details}

\subsection{Choice of gauge}

Since we are calculating the gauge-independent
quantity~$V({\rm min})$,
the gauge may be chosen at will.
The simplest possibility is the $R_\xi$-gauge
with $\xi=1$.
It is not suitable for calculating the effective potential
for arbitrary $\varphi$~(see, e.g.,~\cite{K}),
but when $V({\rm min})$ is extracted from $V(\varphi)$
consistently in powers of $\hbar$ using eq.~\eref{value},
the $R_\xi$-gauge can be used~\cite{KLS}. Hence
the difference between the two-loop contributions to
the effective potential in the L- and $\overline{\rm MS}$-schemes can be
calculated in this gauge.

To be absolutely sure, one could also
just calculate all the connected graphs
in the classical broken minimum, since
then no reference is made to the effective
potential. This amounts to fixing
\beq
\varphi\equiv\sqrt{-m_3^2(\mu)/\lambda_3},
\label{clmin}
\eeq
where $\mu$ is chosen so that $\varphi$ is real,
and adding all the reducible two-loop graphs to the irreducible
ones contributing to the effective potential.
We shall indicate below the differences in
the intermediate stages of the two mentioned
ways of organizing the calculation.

The unshifted Lagrangian needed at the two-loop level
is obtained by expanding eq.~\eref{Sl}
in powers of~$A_i^a$, and
by adding the gauge-fixing and the ghost term.
The gauge fixing term is chosen
as ${\cal L}_\xi=F^aF^a/2\xi a^2$, where
\beq
F^a(x)=\sum_i\Bigl[A_i^a(x)-A_i^a(x-i)\Bigr]+
\frac{1}{2}\xi a g_3\varphi\phi_a(x) .
\label{gf}
\eeq
Gauge fixing is compensated for by
the Faddeev-Popov determinant
$\det [\partial F^a(x)/\partial \theta^b(y)]$, where the $\theta^b(y)$
parametrize the gauge transformations of eq.~\eref{gt} as
\beq
g(x)=\exp[\frac{i}{2}{ag_3}\tau^b\theta^b(x)].
\eeq
With these terms added, the unshifted Lagrangian
is complete.

To get the shifted Lagrangian needed for calculating~$V(\varphi)$,
one replaces $\phi_0$ by $\phi_0+\varphi$. The non-diagonal
terms between $A_i^a$ and $\phi_a$ are cancelled due to eq.~\eref{gf}.
If one is calculating the effective potential,
all the linear terms are neglected.
If one is calculating the value of the path integral
in the broken minimum, the linear
term $\varphi\phi_0[m_3^2(\mu)+\lambda_3\varphi^2]$ vanishes
due to eq.~\eref{clmin},
but the counterterm $\delta m_3^2\varphi\phi_0$ remains.
This enters when reducible two-loop graphs
of the type in Fig.~2 of~\cite{M} are calculated.

\subsection{Feynman rules}

{}From the shifted Lagrangian, one can read the
Feynman rules of the theory. From now on, we take
the gauge parameter equal to unity, $\xi=1$.
The masses of the shifted theory are
\beq
m_T^2\equiv \frac{1}{4} {g_3^2\varphi^2},\quad
m_1^2\equiv m_3^2(\mu)+3\lambda_3\varphi^2,\quad
m_2^2\equiv m_3^2(\mu)+\lambda_3\varphi^2,\quad {m_2'}^2=m_2^2+m_T^2.
\label{masses}
\eeq
If $\varphi$ is chosen according to eq.~\eref{clmin},
the Goldstone boson mass squared $m_2^2$ vanishes.
However, it is useful to keep it in calculations
even in this case,
since this allows one to separate the unphysical
vacuum contributions. Indeed, the vacuum contributions
are obtained by calculating the value of the loop expansion
in the symmetric phase, which means putting
$\varphi\to 0$ in all the expressions, i.e., calculating $V(0)$.

To display the propagators and the vertices
of the theory, we use the notations
\beq
\widetilde{p}_i=\frac{2}{a}\sin\frac{a}{2}p_i, \qquad
\widetilde{p}^2=\sum_i \widetilde{p}_i^2, \qquad
\undertilde{p_i}=\cos\frac{a}{2}p_i. \label{notation}
\eeq
The propagators are then
\bea
\langle{\phi_0(p)}{\phi_0(-p)}\rangle & = &
\frac{1}{\widetilde{p}^2+m_1^2} \quad\quad\quad\,\,
\langle{\phi_a(p)}{\phi_b(-p)}\rangle  =
\frac{\delta_{ab}}{\widetilde{p}^2+{m_2'}^2}
 \nonumber \\
\langle{\overline{c}^a(p)}{c^b(p)}\rangle & = &
-\frac{\delta^{ab}}{\widetilde{p}^2+m_T^2}\quad\quad
\langle{A_i^a(p)}{A_j^b(-p)}\rangle = \delta^{ab}
\frac{\delta_{ij}}{\widetilde{p}^2+m_T^2}\label{propagators}.
\eea
The vertices relevant for the two-loop calculation are as
follows. From~\cite{R} one can read the
two-gluon vertex~[eq.~(14.39) with $1/4a^2\to 1/6a$],
the three-gluon vertex~[eq.~(14.43)],
the four-gluon vertex~[eq.~(14.44) with
$(2/3)(\delta_{AB}\delta_{CD}+\ldots)\to (\delta_{AB}\delta_{CD}+\ldots)$
and $d_{ABC}\to 0$], and the two gluon-ghost vertices~[on pages 212
and 213, with the sign of the $\overline{c}cAA$-vertex changed].
The remaining part of the action reads
\bea
S_{\phi} & = &
\frac{1}{2}\delta m_L^2 (\phi_0^2+\phi_a\phi_a)+
\delta m_L^2\varphi\phi_0 \nonumber \\
& + & \frac{1}{4}{\lambda_3}\Bigl[\phi_0^4+2\phi_0^2\phi_a\phi_a+
\phi_a\phi_a\phi_b\phi_b\Bigr] + \lambda_3 \varphi \phi_0\Bigl[\phi_0^2+
\phi_a\phi_a\Bigr]\nonumber \\
& - & \frac{1}{2}ig_3\delta(p+q+r)\widetilde{(p_i-r_i)}
\biggl[\phi_0(p)A^a_i(q)\phi_a(r)+
\frac{1}{2}\epsilon_{abc}\phi_a(p)A_i^b(q)\phi_c(r)\biggr]\nonumber \\
& + & \frac{1}{8}g_3^2\delta(p+q+r+s)\wideundertilde{(r_i-s_i)}
A_i^a(p)A_i^a(q)\Bigl[
\phi_0(r)\phi_0(s)+\phi_b(r)\phi_b(s)\Bigr]\label{vertices} \\
& + & \frac{1}{4}g_3^2\varphi\delta(p+q+r)\undertilde{r_i}
A_i^a(p)A_i^a(q)\phi_0(r)-\frac{1}{384}a^2g_3^4\varphi^2
A_i^aA_i^aA_i^bA_i^b\nonumber \\
& + & \frac{1}{4}g_3^2\varphi
\overline{c}^ac^b\Bigl[
\delta_{ab}\phi_0+
\epsilon_{acb}\phi_c\Bigr]\nonumber
\nonumber  ,
\eea
where due summations and integrations are implied.
The tree-level part was not displayed,
and the linear counterterm $\delta m_L^2\varphi\phi_0$
is not needed for $V(\varphi)$.
The integration measure is
\beq
\int dp \equiv\int_{-\pi/a}^{\pi/a}\frac{d^3p}{(2 \pi)^3} ,
\eeq
and $\delta(p)$ is a shorthand for $(2\pi)^3\delta_P(p)$,
where $\delta_P(p)$ is periodic with period $2\pi/a$.

\subsection{Integrations}

In the limit $a\to 0$, eq.~\eref{vertices} naturally reproduces
the corresponding part of the action of the theory in
eq.~\eref{Sc}, apart from counterterms.
However, when individual graphs are calculated with finite~$a$,
and the limit $a\to 0$ is taken only after the integrations,
results differ from those in the $\overline{\rm MS}$-scheme.
In this Section we work out
the differences of the one- and two-loop contributions. Mass-independent
vacuum terms of the form $1/a^3$ and $g_3^2/a^2$ are neglected.

Let us start by calculating the one-loop counterterms.
These are known from~\cite{FKRS3},
but we repeat the calculation. In the limit $a\to 0$,
the difference of the one-loop effective potentials
of the L- and $\overline{\rm MS}$-schemes is
\beq
V_L(\varphi)-V_{\overline{\rm MS}}(\varphi)=
\frac{1}{2}\delta m_L^2(\hbar)\varphi^2+
\frac{\hbar}{2}\frac{\Sigma}{4\pi a}(6m_T^2+m_1^2+3m_2^2),
\label{1l}
\eeq
where
\beq
\Sigma=\frac{1}{\pi^2}\int_{-\pi/2}^{\pi/2}d^3x
\frac{1}{\sum_i\sin^2x_i}.\label{sigma}
\eeq
Eq.~\eref{1l} can easily be calculated in a general gauge,
and is seen to be gauge-independent.
Apart from vacuum terms, the two schemes must give
the same result, and hence the
difference in eq.~\eref{1l} must disappear.
Using eq.~\eref{masses} one then sees that
\beq
\delta m_L^2(\hbar)=-\biggl(\frac{3}{2}g_3^2+6\lambda_3\biggr)
\frac{\hbar\Sigma}{4\pi a}.
\label{d31}
\eeq
The mass-dependent vacuum counterterm,
needed to make the vacuum part of the
right-hand side of eq.~\eref{1l}
disappear, is
\beq
\delta V_L(\hbar)=-2m_3^2\frac{\hbar\Sigma}{4\pi a}.
\label{vac31}
\eeq

\begin{figure}[t]
\vspace*{1cm}

\hspace*{2.5cm}
\epsfysize=7cm
\epsffile{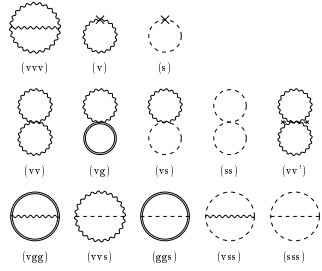}
\begin{center}
\begin{minipage}[t]{15cm}
\renewcommand{\baselinestretch}{1.2}
\caption[a]{\protect
The irreducible two-loop graphs contributing
to the two-loop effective potential in the L-scheme.
Wiggly line is the vector propagator, dashed line is
the scalar propagator, and double line is the ghost propagator.}
\label{graphs}
\end{minipage}
\end{center}
\end{figure}

The two-loop graphs are naturally much more tedious
than the one-loop graphs.
For illustration, we calculate the most complicated of them
in some detail. This is the graph (vvv) in Fig.~\ref{graphs}.
{}From the Feynman rules it follows that
\beq
({\rm vvv})=-\frac{1}{2}g_3^2\int\! dp\,dq\,dr\,\delta(p+q+r) \frac{F(p,q,r)}
{(\widetilde{p}^2+m_T^2)(\widetilde{q}^2+m_T^2)
(\widetilde{r}^2+m_T^2)},
\label{vvv}
\eeq
where
\beq
F(p,q,r)=\sum_{i,j,k}
\Bigl[
\delta_{kj}\undertilde{p_j}\widetilde{(r_i-q_i)}
+\delta_{ik}\undertilde{q_k}\widetilde{(p_j-r_j)}
+\delta_{ji}\undertilde{r_i}\widetilde{(q_k-p_k)}
\Bigr]^2 .
\eeq
Using trigonometric identities, one can express $F(p,q,r)$
in terms of products of the functions $\widetilde{p_i}$,
$\widetilde{q_i}$, and $\widetilde{r_i}$. Utilizing the
symmetry of eq.~\eref{vvv} in exchanges of $p$, $q$, and $r$,
one then gets
\bea
F(p,q,r) & \Rightarrow &
3\Bigl(3-\frac{1}{4}a^2\widetilde{r}^2\Bigr)
\Bigl(2\widetilde{p}^2+2\widetilde{q}^2-\widetilde{r}^2
-a^2\sum_i\widetilde{p_i}^2\widetilde{q_i}^2\Bigr) \nonumber \\
& - &
3\Bigl(3\widetilde{r}^2
-2a^2\sum_i\widetilde{p_i}^2\widetilde{r_i}^2
+\frac{1}{4}a^2\sum_i\widetilde{r_i}^4
+\frac{1}{4}a^4\sum_i\widetilde{p_i}^2\widetilde{q_i}^2\widetilde{r_i}^2
\Bigr). \label{F}
\eea
Next, factors of $m_T^2$ are added and subtracted in eq.~\eref{F},
so that terms of the form $\widetilde{p}^2+m_T^2$
cancel against similar terms in the denominator of eq.~\eref{vvv}.
As a result, the following nine types of integrals remain:

{\bf 1.} There is the integral
\beq
I_1=g_3^2m_T^2 \int\! dp\,dq\;\frac{1}{[\widetilde{p}^2+m_T^2]
[\widetilde{q}^2+m_T^2][\widetilde{(p+q)}^2+m_T^2]}\equiv
g_3^2m_T^2 H_a(m_T,m_T,m_T) . \label{Ha}
\eeq
In~\cite{FKRS3} $H_a(m_T,m_T,m_T)$ was parametrized as
\beq
H_a(m_T,m_T,m_T)=
\frac{1}{16\pi^2}\biggl[\log\frac{2}{a m_T}+\frac{1}{2}+
\zeta+{\cal O}(a)\biggr] ,\label{He}
\eeq
and from~\cite{FKRS1} it is known that
\beq
H_a(m_T,m_T,m_T)=H_c(m_T,m_T,m_T)+\frac{1}{16\pi^2}
\biggl[\log\frac{6}{a\mu}+\zeta\biggr]
+{\cal O}(a) , \label{i1}
\eeq
where $H_c(m_T,m_T,m_T)$ is the finite part of the continuum
limit of $H_a(m_T,m_T,m_T)$ in the $\overline{\rm MS}$-scheme.
The contributions of eq.~\eref{Ha}
to the renormalized two-loop effective potential~$V_2(\varphi)$
in the lattice and $\overline{\rm MS}$ -schemes hence differ
by $(g_3^2m_T^2/16\pi^2)[\log (6/a\mu)+\zeta]$. There is also
a contribution of the form $g_3^2m_T^4a^2H_a(m_T,m_T,m_T)$ from
the diagram (vvv) to $V_2(\varphi)$,
but by eq.~\eref{He} this vanishes in the continuum limit.

{\bf 2.} There is the integral
\beq
I_2=g_3^2\int\! dp\,dq\;\frac{1}{(\widetilde{p}^2+m_T^2)
(\widetilde{q}^2+m_T^2)}\equiv
g_3^2I_a(m_T)I_a(m_T) ,
\eeq
where $I_a(m_T)$ is~\cite{FKRS3}
\beq
I_a(m_T)\equiv\int\! dp\,\frac{1}{\widetilde{p}^2+m_T^2}=
\frac{1}{4\pi}\Biggl[\frac{\Sigma}{a}-m_T
-\xi a m_T^2\Biggr]+{\cal O}(a^2).\label{Ia}
\eeq
Note that the $\xi$ appearing in eq.~\eref{Ia} has
nothing to do with the $\xi$ in eq.~\eref{gf};
the latter has been fixed to unity.
In the $\overline{\rm MS}$-scheme $I_a(m_T)$
is replaced by $I_c(m_T)=-m_T/4\pi$, so that the difference
between the contributions of the two schemes is
\beq
I_a(m_T)I_a(m_T)-I_c(m_T)I_c(m_T)=
\frac{\Sigma^2}{16\pi^2}\frac{1}{a^2}
-\frac{\Sigma}{8\pi^2}\frac{1}{a}m_T
-\frac{\Sigma}{8\pi^2}\xi m_T^2
+{\cal O}(a) . \label{i2}
\eeq

{\bf 3.} The integral
\beq
I_3=g_3^2m_T^2a^2\int\! dp\,dq\;\frac{1}{(\widetilde{p}^2+m_T^2)
(\widetilde{q}^2+m_T^2)}\equiv
g_3^2m_T^2a^2I_a(m_T)I_a(m_T)
\eeq
has no analogue in the $\overline{\rm MS}$-scheme,
but gives by eq.~\eref{i2}
the finite contribution $g_3^2m_T^2\Sigma^2/16\pi^2$ in the L-scheme.

{\bf 4.} The integral
\beq
I_4=g_3^2a^2\int\! dp\,dq\;\frac{1}{\widetilde{p}^2+m_T^2}
=g_3^2I_a(m_T)\frac{1}{a}
\eeq
has no analogue in the $\overline{\rm MS}$-scheme, but gives by eq.~\eref{Ia}
the term $-g_3^2({m_T}{a}^{-1}+\xi m_T^2)/{4\pi}$ in the L-scheme.

{\bf 5.} The integral
\bea
I_5=g_3^2a^4\int\! dp\,dq\;\frac{\sum_i\widetilde{p_i}^2
\widetilde{q_i}^2}{(\widetilde{p}^2+m_T^2)(\widetilde{q}^2+m_T^2)} & = &
\frac{1}{3}g_3^2a^4\int\! dp\,dq\;\frac{\widetilde{p}^2
\widetilde{q}^2}{(\widetilde{p}^2+m_T^2)
(\widetilde{q}^2+m_T^2)} \nonumber \\
& = & \frac{g_3^2}{3a^2}-\frac{2}{3}g_3^2m_T^2aI_a(m_T)+
{\cal O}(a^2)
\eea
has no analogue in the $\overline{\rm MS}$-scheme, but gives by eq.~\eref{Ia}
$-g_3^2m_T^2\Sigma/6\pi$ in the L-scheme.

{\bf 6.} There is the integral
\bea
I_6 & = & g_3^2a^2\int\! dp\,dq\;\frac{\sum_i\widetilde{p_i}^2
\widetilde{(p_i+q_i)}^2}
{[\widetilde{p}^2+m_T^2][\widetilde{q}^2+m_T^2]
[\widetilde{(p+q)}^2+m_T^2]}\nonumber \\
&  = & \frac{g_3^2m_T^2}{16\pi^6}\frac{1}{z^2}
\int_{\pi/2}^{\pi/2}\! d^3xd^3y\;\frac{\sum_i\sin^2{x_i}
\sin^2{(x_i+y_i)}}
{[\sum_i\sin^2{x_i}+z^2]
[\sum_i\sin^2{y_i}+z^2]
[\sum_i\sin^2{(x_i+y_i)}+z^2]},
\eea
where $z^2={a^2m_T^2}/{4}$.
The integral $I_6$ contains
a linear $1/a$-divergence, which can be separated
by adding and subtracting
$g_3^2\alpha I_a(m_T)/a$, where
\beq
\alpha=a^3\int\! dp\frac{\sum_i\widetilde{p_i}^4}
{(\widetilde{p}^2)^2}
=\frac{1}{\pi^3}\int_{-\pi/2}^{\pi/2}d^3x
\frac{\sum_i\sin^4x_i}{(\sum_i\sin^2x_i)^2}.\label{alpha}
\eeq
In the rest of the integral, writing the propagators in the form
\beq
\frac{1}{\sum_i\sin^2{x_i}+z^2}=
-\frac{z^2}{(\sum_i\sin^2{x_i}+z^2)(\sum_i\sin^2{x_i})}+
\frac{1}{\sum_i\sin^2{x_i}}
\label{fraction}.
\eeq
allows one to separate the vacuum part $g_3^2/a^2$,
the finite part $g_3^2m_T^2$,
and the part vanishing with $a$.
Neglecting the vacuum part, the result is
\beq
I_6=\frac{g_3^2 \alpha}{a} I_a(m_T)-\frac{g_3^2m_T^2}{4\pi^2}(\delta+\rho)
+{\cal O}(a),
\eeq
where
\beq
\delta=\frac{1}{2\pi^4}\int_{-\pi/2}^{\pi/2}d^3xd^3y
\frac{\sum_i\sin^2x_i\sin^2(x_i+y_i)}
{(\sum_i\sin^2x_i)^2\sum_i\sin^2(x_i+y_i)\sum_i\sin^2y_i},
\eeq
\beq
\rho=\frac{1}{4\pi^4}\int_{-\pi/2}^{\pi/2}d^3xd^3y
\Biggl\{\frac{\sum_i\sin^2x_i\sin^2(x_i+y_i)}
{\sum_i\sin^2x_i\sum_i\sin^2(x_i+y_i)}-
\frac{\sum_i\sin^4x_i}
{(\sum_i\sin^2x_i)^2}\Biggr\}
\frac{1}{(\sum_i\sin^2y_i)^2}.
\eeq

{\bf 7.} In the integral
\beq
I_7 = g_3^2m_T^2a^4\int\! dp\,dq\;\frac{\sum_i\widetilde{p_i}^2
\widetilde{(p_i+q_i)}^2}
{[\widetilde{p}^2+m_T^2][\widetilde{q}^2+m_T^2]
[\widetilde{(p+q)}^2+m_T^2]},\nonumber
\eeq
the mass terms in the propagators give contributions of higher
order in~$a$. Hence $I_7$ has in the limit $a\to 0$
the value $I_7=g_3^2m_T^2\kappa_1/\pi^2$, where
\beq
\kappa_1=\frac{1}{4\pi^4}\int_{-\pi/2}^{\pi/2}d^3xd^3y
\frac{\sum_i\sin^2x_i\sin^2(x_i+y_i)}
{\sum_i\sin^2x_i\sum_i\sin^2(x_i+y_i)\sum_i\sin^2y_i}.
\eeq

{\bf 8.} The integral
\beq
I_8=g_3^2a^2\int\! dp\,dq\;\frac{\sum_i\widetilde{(p_i+q_i)}^4}
{[\widetilde{p}^2+m_T^2][\widetilde{q}^2+m_T^2]
[\widetilde{(p+q)}^2+m_T^2]}
\eeq
can be handled exactly as $I_6$. The result is
\beq
I_8=2\frac{g_3^2\alpha}{a} I_a(m_T)-\frac{g_3^2m_T^2}{4\pi^2}(\kappa_2+
\kappa_3)
+{\cal O}(a),
\eeq
where
\beq
\kappa_2=\frac{1}{4\pi^4}\int_{-\pi/2}^{\pi/2}d^3xd^3y
\frac{\sum_i\sin^4x_i}
{(\sum_i\sin^2x_i)^2\sum_i\sin^2(x_i+y_i)\sum_i\sin^2y_i},
\eeq
\beq
\kappa_3=\frac{1}{2\pi^4}\int_{-\pi/2}^{\pi/2}d^3xd^3y
\Biggl\{\frac{1}
{\sum_i\sin^2x_i\sum_i\sin^2(x_i+y_i)}-
\frac{1}
{(\sum_i\sin^2x_i)^2}\Biggr\}
\frac{\sum_i\sin^4x_i}{(\sum_i\sin^2y_i)^2}.
\eeq

{\bf 9.} The integral
\beq
I_9=g_3^2a^4\int\! dp\,dq\;\frac{\sum_i
\widetilde{p_i}^2\widetilde{q_i}^2\widetilde{(p_i+q_i)}^2}
{[\widetilde{p}^2+m_T^2][\widetilde{q}^2+m_T^2]
[\widetilde{(p+q)}^2+m_T^2]}
\eeq
can be simplified with the method of eq.~\eref{fraction}, giving
$I_9=-3g_3^2m_T^2\kappa_4/4\pi^2$, where
\beq
\kappa_4=\frac{1}{\pi^4}\int_{-\pi/2}^{\pi/2}d^3xd^3y
\frac{\sum_i\sin^2x_i\sin^2(x_i+y_i)\sin^2y_i}
{(\sum_i\sin^2x_i)^2\sum_i\sin^2(x_i+y_i)\sum_i\sin^2y_i}.
\eeq
This completes the enumeration of the integrals
that appear in the graph (vvv).

In addition to the continuum contributions taken into
account when discussing the integrals $I_1$ and $I_2$
in eqs.~\eref{i1} and~\eref{i2}, there are
extra continuum contributions in the $\overline{\rm MS}$-scheme. Namely,
the graphs (vvv) and (vvs) contain a part where the trace
of the metric tensor $\delta_{ii}=3-2\epsilon$ multiplies
the function $H_c\propto 1/\epsilon$. The finite
contributions arising from the products of $\epsilon$
and $1/\epsilon$ are
\beq
{\rm (vvv)}\Rightarrow
-\frac{\hbar^2}{16\pi^2}\frac{9}{16}g_3^4\varphi^2,
\qquad{\rm (vvs)}\Rightarrow
\frac{\hbar^2}{16\pi^2}\frac{3}{32}g_3^4\varphi^2.
\label{extra}
\eeq
Naturally, this kind of contributions do not arise on lattice.

When all the numerical factors are taken into account,
the integrals $I_1$--$I_9$ and the extra continuum contributions
in eq.~\eref{extra} finally yield for the
difference of the L and $\overline{\rm MS}$-schemes from
the graph (vvv) the value
\bea
({\rm vvv}) & \Rightarrow & g_3^2(18\Sigma-3\pi-9\pi\alpha)\frac{m_T}{a}+
\frac{9}{4}{g_3^4\varphi^2}
\biggl(\log\frac{6}{a\mu}+\zeta\biggr)  \nonumber \\
& + &
\frac{3}{8}g_3^4\varphi^2
\biggl(\frac{3}{2}-\frac{5}{4}\Sigma^2+\frac{\pi}{3}\Sigma
-4\delta-4\rho+4\kappa_1-\kappa_2-\kappa_3-3\kappa_4\biggr).
\label{first}
\eea
Here a common factor $\hbar^2/16\pi^2$ has been neglected. In addition,
terms proportional to $\xi$ are not shown explicitly,
since one can see from the above that a term of the from $\xi m^2$ is
always accompanied with the term $m/a$. In the end, the $1/a$-terms
will cancel, so that the $\xi$-terms also cancel.

To conclude this Section, we list the differences of the L
and $\overline{\rm MS}$-schemes to $V(\varphi)$ from the rest of the
irreducible two-loop graphs,
shown in Fig.~\ref{graphs}. Again the factor $\hbar^2/16\pi^2$,
and all terms proportional to $\xi$,
are neglected. The graph (s) arises from the
one-loop mass-counterterm in eq.~\eref{d31},
and (v) arises from the gluon-gluon vertex induced
by the Haar measure. The graph (vv') arises from the $\varphi^2A^4$-vertex
in eq.~\eref{vertices}:
\FL
\beq
({\rm v})\Rightarrow -3\pi g_3^2\frac{m_T}{a}
\eeq
\FL
\beq
({\rm s})\Rightarrow \frac{3}{4}(g_3^2+4\lambda_3)
\Sigma\biggl(\frac{m_1}{a}+3\frac{m_2'}{a}\biggr)
\eeq
\FL
\beq
({\rm vv})\Rightarrow g_3^2(14\pi-18\Sigma)\frac{m_T}{a}+
\frac{7}{8}{g_3^4\varphi^2}\Sigma^2
\eeq
\FL
\beq
({\rm vg})\Rightarrow -2\pi g_3^2\frac{m_T}{a}-
\frac{1}{8}g_3^4\varphi^2{\Sigma^2}
\eeq
\FL
\beq
({\rm vs})\Rightarrow -\frac{3}{2}g_3^2(3\Sigma-2\pi)
\frac{m_T}{a}-
\frac{9}{8}g_3^2\Sigma\biggl(\frac{m_1}{a}+3\frac{m_2'}{a}\biggr)+
\frac{3}{4}g_3^2m_3^2\Sigma^2+
\frac{9}{64}(g_3^4+8\lambda_3g_3^2)\varphi^2\Sigma^2 \label{vs}
\eeq
\FL
\beq
({\rm ss})\Rightarrow -3\lambda_3\Sigma
\biggl(\frac{m_1}{a}+3\frac{m_2'}{a}\biggr)
\eeq
\FL
\beq
({\rm vv'})\Rightarrow -\frac{15}{128}g_3^4\varphi^2\Sigma^2
\eeq
\FL
\beq
({\rm vgg})\Rightarrow 3g_3^2(\pi\alpha-\Sigma)\frac{m_T}{a}+
\frac{3}{4}g_3^4\varphi^2(\delta+\rho)-
\frac{3}{8}g_3^4\varphi^2\biggl(\log\frac{6}{a\mu}+\zeta\biggr)
\eeq
\FL
\beq
({\rm vvs})\Rightarrow \frac{3}{64}g_3^4\varphi^2(\Sigma^2-2)
-\frac{9}{16}g_3^4\varphi^2\biggl(\log\frac{6}{a\mu}+\zeta\biggr)
\eeq
\FL
\beq
({\rm ggs})\Rightarrow -
\frac{3}{32}g_3^4\varphi^2\biggl(\log\frac{6}{a\mu}+\zeta\biggr)
\eeq
\FL
\bea
({\rm vss}) & \Rightarrow  & 3g_3^2(\Sigma-\pi\alpha)\frac{m_T}{a}+
\frac{3}{8}g_3^2\Sigma(\frac{m_1}{a}+3\frac{m_2'}{a})-
\frac{3}{16}
g_3^4\varphi^2(3\delta+4\rho)
-\frac{9}{2} \lambda_3g_3^2\varphi^2\delta
\nonumber \\
& + & \frac{3}{8}
(g_3^4+12\lambda_3g_3^2)\varphi^2
\biggl(\log\frac{6}{a\mu}+\zeta\biggr)
+3g_3^2m_3^2\biggl(\log\frac{6}{a\mu}+\zeta-\delta\biggr) \label{vss}
\eea
\FL
\beq
({\rm sss})\Rightarrow -6\lambda_3^2\varphi^2
\biggl(\log\frac{6}{a\mu}+\zeta\biggr).
\label{last}
\eeq

\section{Results at two-loop level}
\label{results}

We are now ready to sum together the
differences of the L- and $\overline{\rm MS}$-schemes from
all the two-loop graphs. First, let us note that
the reducible two-loop graphs of the type in Fig.~2
of~\cite{M}, needed when calculating the value of the path integral
in the broken minimum, give exactly the same result in the L-
and $\overline{\rm MS}$-schemes,
and hence do not affect $\delta m_L^2(\hbar^2)$.
Only the irreducible graphs of Fig.~\ref{graphs} are significant.
Second, when the eqs.~\eref{first}--\eref{last}
are summed together, all the terms proportional
to $m_1/a$ and $m'_2/a$ cancel. Apart from
vacuum terms, this leaves the result
\bea
V_L(\varphi)-V_{\overline{\rm MS}}(\varphi)& = &
\frac{1}{2}\delta m_L^2(\hbar^2)\varphi^2+
\frac{\hbar^2}{16\pi^2}9\pi g_3^2
\biggl(1-\frac{\Sigma}{2\pi}-\alpha\biggr)\frac{m_T}{a} \nonumber \\
& + & \frac{\hbar^2}{16\pi^2}\frac{\varphi^2}{2}\biggl[
\biggl(\frac{51}{16}g_3^4+9\lambda_3 g_3^2-12\lambda_3^2\biggr)
\biggl(\log\frac{6}{a\mu}+\zeta\biggr)+
9\lambda_3 g_3^2\biggl(\frac{1}{4}\Sigma^2-\delta\biggr) \nonumber \\
& + &
\frac{3}{4}g_3^4\biggl(\frac{15}{16}\Sigma^2+
\frac{\pi}{3}\Sigma+\frac{5}{4}-\frac{7}{2}\delta
-4\rho+4\kappa_1-\kappa_2-\kappa_3-3\kappa_4\biggr)\biggr]\,.
\label{del}
\eea
{}From the identity
\beq
0=\int_{-\pi/2}^{\pi/2}d^3x\frac{d}{dx_1}\frac{\sin x_1\cos x_1}
{\sin^2x_1+\sin^2x_2+\sin^2x_3},
\eeq
it follows that $\alpha=1-\Sigma/2\pi$,
and hence the $m_T/a$-terms cancel.
Since eq.~\eref{del} must vanish, we finally get
\bea
\delta m_L^2(\hbar^2) & = &
-\frac{\hbar^2}{16\pi^2}\biggl[
\biggl(\frac{51}{16}g_3^4+9\lambda_3 g_3^2-12\lambda_3^2\biggr)
\biggl(\log\frac{6}{a\mu}+\zeta\biggr)
+9\lambda_3 g_3^2\biggl(\frac{1}{4}\Sigma^2-\delta\biggr)\nonumber \\
& & +\frac{3}{4}g_3^4\biggl(\frac{15}{16}\Sigma^2+
\frac{\pi}{3}\Sigma+\frac{5}{4}-\frac{7}{2}\delta
-4\rho+4\kappa_1-\kappa_2-\kappa_3-3\kappa_4\biggr)\biggr].
\label{d32}
\eea

{}From eqs.~\eref{vs} and~\eref{vss}, one can also read
the mass-dependent two-loop vacuum counterterm needed on lattice,
in order to make the renormalized mass-dependent vacuum parts of the
effective potentials the same in the two schemes:
\beq
\delta V_L(\hbar^2)=
-\frac{\hbar^2}{16\pi^2}3g_3^2m_3^2(\mu)
\biggl(\log\frac{6}{a\mu}+\zeta+\frac{\Sigma^2}{4}-\delta\biggr).
\label{vac32}
\eeq
Note that the $\mu$-dependence of eq.~\eref{vac32}
reproduces that of eq.~\eref{MSV}.

Eq.~\eref{d32} contains
eight pure numbers, $\zeta$, $\Sigma$, $\delta$, $\rho$,
$\kappa_1$,$\kappa_2$,$\kappa_3$, and $\kappa_4$.
The parameter $\Sigma$, defined in eq.~\eref{sigma},
is known analytically~\cite{FKRS3}, and its numerical
value is $\Sigma\approx3.176$. From the identity
\beq
0=\int_{-\pi/2}^{\pi/2}d^3xd^3y\frac{d}{dx_1}\frac{\sin x_1\cos x_1}
{[\sum_i\sin^2x_i+m^2][\sum_i\sin^2(x_i+y_i)+m^2]
[\sum_i\sin^2y_i+m^2]},
\eeq
it follows that $\kappa_2=\Sigma^2/4-\delta/2-1/4$.
This still leaves six parameters to be calculated numerically.
In~\cite{FKRS3} the values $\zeta\approx0.09$, $\delta\approx1.94$, and
$\rho\approx-0.314$ were given. We have calculated that
$\kappa_1\approx0.958$, $\kappa_3\approx0.751$,
and $\kappa_4\approx1.20$.
The accuracy of these numbers could probably be
considerably improved with the techniques
of~\cite{LW} adapted to three dimensions,
but we have not attempted to do so.
Numerically we then have that
\beq
\delta m_L^2(\hbar^2) \approx
-\frac{\hbar^2}{16\pi^2}\biggr[
\biggl(\frac{51}{16}g_3^4+9\lambda_3 g_3^2-12\lambda_3^2\biggr)
\biggl(\log\frac{6}{a\mu}+0.09\biggr)+
5.0g_3^4+5.2\lambda_3 g_3^2\biggr]. \nonumber
\eeq

There are a few way of checking parts of the
analytic result in eq.~\eref{d32}. First,
the cancellation of $1/a$-divergences indicating the
renormalizability of the theory is a non-trivial check,
since such terms arise from most of the graphs.
Second, from eqs.~\eref{mr} and~\eref{d32} one sees that
the $\mu$-dependence cancels in eq.~\eref{CTs}, as it should.
Third, in~\cite{FKRS3} the
mass counterterm in the L-scheme was
determined by a combination of
analytical and lattice Monte Carlo methods.
The parts proportional to $\lambda_3g_3^2$ and $\lambda_3^2$
in $\delta m_L^2(\hbar^2)$ were determined analytically, and
they agree with eq.~\eref{d32}. The part
proportional to $g_3^4$ was determined by lattice
Monte Carlo methods;
for the SU(2)~+ fundamental Higgs theory, the
coefficient of~$g_3^4$
was parametrized with the number $\eta_0=2.12(7)$, and
for the SU(2) + fundamental Higgs + adjoint Higgs theory,
with the number $\eta=2.18(6)$. Eq.~\eref{d32} implies
for $\eta_0$ the value~$2.01$, and eq.~\eref{dd32} for
$\eta$ the value~$1.96$. The systematical error in the
determination of $\eta$ in~\cite{FKRS3} is larger than
in the determination of $\eta_0$, since for the former
it was assumed that $\delta m_D^2(\hbar^2)$, defined
as the sum of eqs.~\eref{dD2} and~\eref{mact}, is negligible.
We conclude that the agreement between our analytical
result and lattice Monte Carlo simulations is good.

\section{Conclusions}
\label{concl}

In this paper, the exact relations between lattice and continuum
regularization schemes in 3D super-renormalizable SU(2) and U(1) gauge
theories with Higgs fields in fundamental and adjoint representations
have been calculated. These relations are needed when results from
lattice simulations are related to continuum observables.
The general structure of the calculated mass counter\-terms
is that, in addition to linear $1/a$-terms and logarithmic $\log a$-terms,
there are two-loop constant terms proportional to $g_3^4$
and $\lambda_3g_3^2$. Here $\lambda_3$ denotes the self-coupling of the
relevant scalar field. Numerically, the $g_3^4$-terms are rather large.
The $g_3^4$-terms are especially significant for the SU(2) gauge theory
with a Higgs field in the adjoint representation, since then
the ``dominant'' logarithmic term $g_3^4\log a$ vanishes. The
results obtained have significance for numerical simulations of gauge
theories at high enough temperatures, so that the theories undergo
dimensional reduction into an effective 3D theory.
In particular, the results are important for numerical simulations of the
cosmological electroweak phase transition.

\section*{Acknowledgements}

I am most grateful to K.~Kajantie and
M.~Shaposhnikov for discussions and advice.

\appendix

\section*{}

In this Appendix, results for mass counterterms
in the lattice regularization scheme are presented
for a number of SU(2) and U(1) gauge theories,
with Higgs fields in fundamental and adjoint representations.

\subsection{SU(2) + adjoint Higgs}

The Lagrangian for the SU(2) + adjoint Higgs theory
consists of the standard plaquette action for gauge
bosons on the first row of eq.~\eref{Sl},
supplemented by the terms
\bea
{\cal L}_{A_0} & = &
\frac{1}{a^2}\sum_i\biggl[
\frac{1}{2}{\rm Tr}A_0(x)A_0(x)-
\frac{1}{2}{\rm Tr}A_0(x+i)U_i^{-1}(x)A_0(x)U_i(x)\biggr] \nonumber \\
& + &\frac{1}{2}m_D^2A_0^aA_0^a
+\frac{1}{4}\lambda_A(A_0^aA_0^a)^2 .
\label{Sadj}
\eea
Here $A_0=A_0^a\tau^a$. The matrix $A_0$ transforms
in gauge transformations as $A_0(x)\to A'_0(x)=g(x)A_0(x)g^{-1}(x)$.
The effective potential is calculated by shifting $A_0^3$ to
be $A_0^3+\alpha$, and the gauge fixing condition is chosen in complete
analogy with eq.~\eref{gf}.
The $A_i^3$- and $c^3$-fields remain massless despite the shift,
and the $A_i^1$-, $A_i^2$-, $c^1$-, and $c^2$-fields get the mass
squared $\hat{m}_T^2=g_3^2\alpha^2$.
The mass squared of $A_0^3$ is $m_D^2+3\lambda_A\alpha^2$,
and that of $A_0^1$ and $A_0^2$ is $m_D^2+\lambda_A\alpha^2+g_3^2\alpha^2$.

The two-loop graphs to be calculated in the SU(2) + adjoint Higgs
theory are the same as those in Fig.~\ref{graphs}, with the $\Phi$-field
replaced by the $A_0$-field, and the vertices corrected appropriately.
As in eq.~\eref{extra}, there are extra continuum
contributions in the $\overline{\rm MS}$-scheme from
the diagrams (vvv) and (vvs).
However, the absolute value of both contributions
is $3g_3^4\alpha^2/32\pi^2$, and the signs are different,
so that these terms cancel.

The one-loop mass-counterterm resulting from eq.~\eref{Sadj} is
\beq
\delta m_D^2(\hbar)=-(4g_3^2+5\lambda_A)\frac{\hbar\Sigma}{4\pi a}
\label{dD1},
\eeq
and the two-loop counterterm is
\bea
\delta m_D^2(\hbar^2) & = &
-\frac{\hbar^2}{16\pi^2}\biggl[
\biggl(20\lambda_A g_3^2-10\lambda_A^2\biggr)
\biggl(\log\frac{6}{a\mu}+\zeta\biggr)
+20\lambda_A g_3^2\biggl(\frac{1}{4}\Sigma^2-\delta\biggr)
\nonumber \\
& & +2g_3^4\biggl(\frac{5}{4}\Sigma^2+
\frac{\pi}{3}\Sigma-6\delta
-6\rho+4\kappa_1-\kappa_2-\kappa_3-3\kappa_4\biggr)\biggr].
\label{dD2}
\eea
Using the numerical values given in Sec.~\ref{results},
$\delta m_D^2(\hbar^2)$ has the approximate value
\beq
\delta m_D^2(\hbar^2)\approx
-\frac{\hbar^2}{16\pi^2}\biggl[
\biggl(20\lambda_A g_3^2-10\lambda_A^2\biggr)
\biggl(\log\frac{6}{a\mu}+0.09\biggr)
+8.7g_3^4+11.6\lambda_A g_3^2\biggr].
\label{mdnu}
\eeq
Note that the constant term proportional to $g_3^4$ is
numerically rather large.
If the coupling constant $\lambda_A$ is very small, as is the case in
the context of the high-temperature electroweak theory, the $g_3^4$-term
gives the dominant contribution in eq.~\eref{mdnu} for
moderate~$a$, since the coefficient of the logarithmic term is vanishing.

The vacuum counterterms, determined analogously to
eqs.~\eref{vac31} and~\eref{vac32}, are
\beq
\delta V_L(\hbar)=-\frac{3}{2}m_D^2\frac{\hbar\Sigma}{4\pi a},
\label{vacD1}
\eeq
\beq
\delta V_L(\hbar^2)=-\frac{\hbar^2}{16\pi^2}6g_3^2m_D^2
\biggl(\log\frac{6}{a\mu}+\zeta+\frac{\Sigma^2}{4}-\delta\biggr).
\label{vacD2}
\eeq

\subsection{SU(2) + fundamental Higgs + adjoint Higgs}

The SU(2) + fundamental Higgs + adjoint Higgs theory
consists of the sum of eqs.~\eref{Sl} and~\eref{Sadj},
together with the interaction term
\beq
{\cal L}_{i}=\frac{1}{2}h_3{\rm Tr}\Phi^\dagger\Phi A_0^aA_0^a .
\eeq
The one-loop
mass counterterm $\delta m_L^2(\hbar)$ is corrected from
the value of eq.~\eref{d31} by the amount
\beq
\Delta[\delta m_L^2(\hbar)]=-3h_3\frac{\hbar\Sigma}{4\pi a},
\eeq
and $\delta m_L^2(\hbar^2)$ is corrected from
eq.~\eref{d32} by the amount
\bea
\Delta[\delta m_L^2(\hbar^2)] & = &
-\frac{\hbar^2}{16\pi^2}\biggl[
\biggl(-\frac{3}{4}g_3^4+12h_3g_3^2-6h_3^2\biggr)
\biggl(\log\frac{6}{a\mu}+\zeta\biggr) \nonumber \\
&  &\hspace*{2.2cm} +12h_3g_3^2\biggl(\frac{\Sigma^2}{4}-\delta\biggr)
-3g_3^4\rho\biggr].
\label{dd32}
\eea
The adjoint mass counterterm
$\delta m_D^2(\hbar)$ of eq.~\eref{dD1} is corrected by
\beq
\Delta[\delta m_D^2(\hbar)]=-4h_3\frac{\hbar\Sigma}{4\pi a},
\eeq
and $\delta m_D^2(\hbar^2)$ of eq.~\eref{dD2} is corrected by
\bea
\Delta[\delta m_D^2(\hbar^2)] & = &
-\frac{\hbar^2}{16\pi^2}\biggl[
\biggl(-g_3^4+6h_3g_3^2-8h_3^2\biggr)
\biggl(\log\frac{6}{a\mu}+\zeta\biggr) \nonumber \\
& & \hspace*{2.2cm}+6h_3g_3^2\biggl(\frac{\Sigma^2}{4}-\delta\biggr)
-4g_3^4\rho\biggr].
\label{mact}
\eea
There are no extra contributions of the type
in eq.~\eref{extra} from the coupling constant~$h_3$.
The mass-dependent vacuum counterterm of the
SU(2) + fundamental Higgs + adjoint Higgs theory
is the sum of eqs.~\eref{vac31}, \eref{vac32},
\eref{vacD1}, and \eref{vacD2}.
Note that for the leading order approximation $h_3=g_3^2/4$ of dimensional
reduction, the logarithmic term in eq.~\eref{mact} vanishes.

\subsection{U(1) + fundamental Higgs}

The Lagrangian for the U(1) + fundamental Higgs theory is, in analogy
with eq.~\eref{Sl},
\bea
{\cal L}_L & = & \frac{1}{a^4 e_3^2}\sum_{i<j}
\biggl\{1-\frac{1}{2}\Bigl[P_{ij}(x)+P^{*}_{ij}(x)\Bigr]\biggr\} \nonumber \\
& + & \frac{2}{a^2}\sum_i\biggl\{
\Phi^*(x)\Phi(x)-\frac{1}{2}\Bigl[
\Phi^*(x)U_i(x)\Phi(x+i)+{\rm c.c.}\Bigr]\biggr\}  \label{USl} \\
& + & m_L^2\Phi^*\Phi+\lambda_3\Bigl(\Phi^*\Phi\Bigr)^2,
\nonumber
\eea
where $U_i(x)=\exp[iae_3A_i(x)]$,
$P_{ij}(x)=U_i(x)U_j(x+i)U_i^{*}(x+j)U_j^{*}(x)$,
and $\Phi=(\phi_0+i\phi_1)/\sqrt{2}$.
A one-loop calculation produces the mass-counterterm
\beq
\delta m_L^2(\hbar) = -(2e_3^2+4\lambda_3)\frac{\hbar\Sigma}{4\pi a},
\label{Ud31}
\eeq
and a two-loop calculation yields
\bea
\delta m_L^2(\hbar^2) & = &
-\frac{\hbar^2}{16\pi^2}\biggl[
\biggl(-4e_3^4+8\lambda_3e_3^2-8\lambda_3^2\biggr)
\biggl(\log\frac{6}{a\mu}+\zeta\biggr)
+8\lambda_3e_3^2\biggl(\frac{1}{4}\Sigma^2-\delta\biggr)
\nonumber \\
& & +e_3^4\biggl(\frac{1}{4}\Sigma^2+
\frac{8\pi}{3}\Sigma-1-2\delta-4\rho\biggr)\biggr].
\label{Ud32}
\eea
Numerically, eq.~\eref{Ud32} gives
\beq
\delta m_L^2(\hbar^2)  \approx
-\frac{\hbar^2}{16\pi^2}\biggl[
\biggl(-4e_3^4+8\lambda_3e_3^2-8\lambda_3^2\biggr)
\biggl(\log\frac{6}{a\mu}+0.09\biggr)
+25.5e_3^4+4.6\lambda_3e_3^2\biggr]
\eeq
There is again one extra continuum contribution
of the type in eq.~\eref{extra}, present in eq.~\eref{Ud32}:
\beq
{\rm (vvs)}\Rightarrow \frac{\hbar^2}{16\pi^2}\frac{1}{2}e_3^4\varphi^2.
\eeq
The vacuum counterterms of the U(1) + fundamental Higgs theory are
\beq
\delta V_L(\hbar)=-m_3^2\frac{\hbar\Sigma}{4\pi a},
\label{Uvac31}
\eeq
\beq
\delta V_L(\hbar^2)=
-\frac{\hbar^2}{16\pi^2}2e_3^2m_3^2
\biggl(\log\frac{6}{a\mu}+\zeta+\frac{\Sigma^2}{4}-\delta\biggr).
\label{Uvac32}
\eeq

\subsection{U(1) + adjoint Higgs}

The U(1) + adjoint Higgs theory is very simple,
since the $A_0$- and $A_i$-fields do not interact.
The results for this theory have been given in \cite{FKRS3}, but
to fix the notation, we restate the results. The $A_i$-part of the
theory is the first row of eq.~\eref{USl}, and the $A_0$-part is
\beq
{\cal L}_{A_0}=
\frac{1}{a^2}\Bigl[A_0(x)A_0(x)-A_0(x)A_0(x+i)\Bigr]+
\frac{1}{2}m_D^2A_0^2+\frac{1}{4}\lambda_A A_0^4.
\label{USadj}
\eeq
The mass counterterms are
\beq
\delta m_D^2(\hbar)=-3\lambda_A\frac{\hbar\Sigma}{4\pi a},
\label{UdD1}
\eeq
\beq
\delta m_D^2(\hbar^2) =
\frac{\hbar^2}{16\pi^2}\biggl[
6\lambda_A^2\biggl(\log\frac{6}{a\mu}+\zeta\biggr)\biggr] .
\label{UdD2}
\eeq
At one loop there is the vacuum counterterm
\beq
\delta V_L(\hbar)=
-\frac{1}{2}m_D^2\frac{\hbar\Sigma}{4\pi a},
\label{UvacD1}
\eeq
but there is no such term at two-loop order.
There are no continuum contributions of the type in eq.~\eref{extra}.

\subsection{U(1) + fundamental Higgs + adjoint Higgs}

The U(1) + fundamental Higgs + adjoint Higgs theory
consists of the sum of eqs.~\eref{USl} and~\eref{USadj},
together with the interaction term
\beq
{\cal L}_{i}=h_3\Phi^*\Phi A_0^2 .
\label{UfA}
\eeq
The one-loop mass counterterm
$\delta m_L^2(\hbar)$ is corrected from
the value of eq.~\eref{Ud31} by the amount
\beq
\Delta[\delta m_L^2(\hbar)]=-h_3\frac{\hbar\Sigma}{4\pi a},
\eeq
and the two-loop counterterm $\delta m_L^2(\hbar^2)$ is corrected from
eq.~\eref{Ud32} by the amount
\beq
\Delta[\delta m_L^2(\hbar^2)]=
\frac{\hbar^2}{16\pi^2}\biggl[
2h_3^2\biggl(\log\frac{6}{a\mu}+\zeta\biggr)\biggr].
\eeq
The adjoint mass counterterm
$\delta m_D^2(\hbar)$ of eq.~\eref{UdD1} is corrected by
\beq
\Delta[\delta m_D^2(\hbar)]=-2h_3\frac{\hbar\Sigma}{4\pi a},
\eeq
and $\delta m_D^2(\hbar^2)$ of eq.~\eref{UdD2} is corrected by
\beq
\Delta[\delta m_D^2(\hbar^2)]=
-\frac{\hbar^2}{16\pi^2}\biggl[
\biggl(4h_3e_3^2-4h_3^2\biggr)
\biggl(\log\frac{6}{a\mu}+\zeta\biggr)
+4h_3e_3^2\biggl(\frac{\Sigma^2}{4}-\delta\biggr)\biggr].
\eeq
There are no extra continuum contributions of the type
in eq.~\eref{extra} from eq.~\eref{UfA}.
The mass-dependent vacuum counterterm of this theory
is the sum of eqs.~\eref{Uvac31}, \eref{Uvac32}, and
\eref{UvacD1}.

\end{document}